\documentclass[revtex4]{emulateapj}

\usepackage{amssymb}
\usepackage{amsmath}
\usepackage[]{graphicx}
\usepackage{enumerate}
\usepackage{subeqnarray}
\usepackage{cases}
\usepackage{mathrsfs,amssymb}

\tightenlines

\begin{document}

\title{Compact dusty clouds and efficient H$_2$ formation in diffuse ISM}

\author{A. V. Ivlev$^1$, A. Burkert$^{1,2}$, A. Vasyunin$^{1,3}$, P. Caselli$^1$}
\email[e-mail:~]{ivlev@mpe.mpg.de} \affiliation{$^1$Max-Planck-Institut f\"ur Extraterrestrische Physik, 85748 Garching,
Germany } \affiliation{$^2$University Observatory Munich, Scheinerstr.~1, 81679 Munich, Germany} \affiliation{$^3$Ural
Federal University, Ekaterinburg, Russia}

\begin{abstract}
The formation of compact dusty clouds in diffuse interstellar medium (ISM) has been recently proposed and studied by
\citet[][]{Tsytovich2014}. In the present paper, an effect of the clouds on the rate of H$\to$H$_2$ transition in the ISM is
examined. We discuss the mechanisms leading to the formation of the clouds and the creation of gaseous clumps around them,
analyze the main processes determining the efficiency of the recombination of atomic hydrogen on dust grains, and
estimate the expected enhancement of the global H$_2$ formation due to the presence of the clouds. In conclusion, we argue
that the compact dusty clouds provide optimum conditions for the atomic recombination in diffuse ISM, and point out some
astrophysical implications of the resulting H$_2$ formation enhancement.
\end{abstract}


\maketitle

\section{Introduction}
\label{intro}

The formation of molecules at the surface of dust grains is a ubiquitous process occurring in various regions of
interstellar medium (ISM). A prominent example is the formation of molecular hydrogen in diffuse ISM, a process of
fundamental importance in astrophysics. Rate and timescales of H$_2$ formation in the diffuse ISM is a matter of debate.
While classical estimates of the formation timescale made by \citet[][]{Hollenbach1971b} give a value of
$\sim3\times10^7$~yr (for the gas density of $\sim10^2$~cm$^{-3}$), there is a strong evidence that this process should
occur significantly faster \citep[e.g.,][]{Hartmann2001,MacLow2004,Liszt2007,MacLow2017}.

The estimates for the reaction rate are based on a ``natural'' (and usually implicit) assumption that the ISM is homogeneous
at (very) small spatial scales. However, there is a growing observational evidence that the gas density in diffuse ISM can
vary substantially at scales below $\sim1$~pc, containing gaseous clumps of sizes down to $\sim1$~AU
\citep[][]{Heiles1997,Crawford2002,Hartquist2003,Heiles2007,Stanimirovic2010,Dutta2014}. The importance of dense clumps in
diffuse clouds was also discussed by \citet[][]{CecchiPestellini2000} and \citet[][]{LePetit2004} for the interpretation of
the large H$_{3}^{+}$ column densities measured in diffuse clouds.

\begin{figure}\centering
\includegraphics[width=.9\columnwidth,clip,trim=0.cm 5.cm 0.cm 5.cm,]{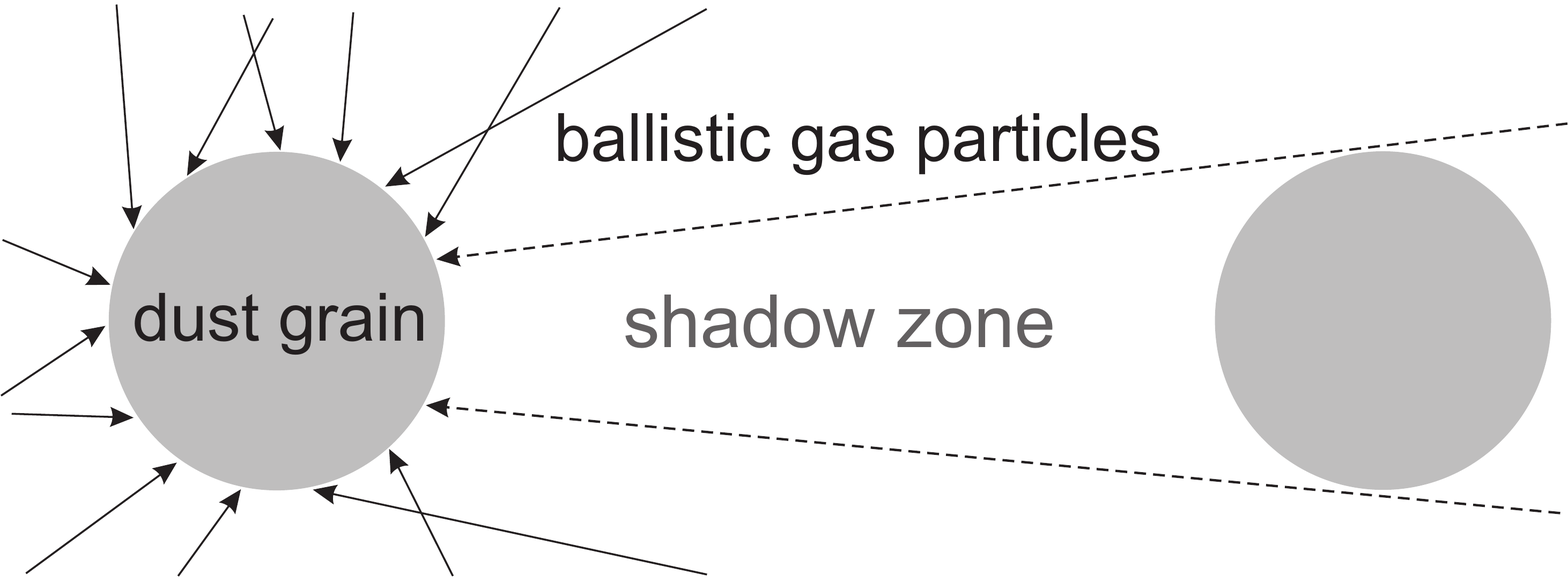}
\caption{Shadowing attraction between dust grains (not to scale). The mean free path of gas
particles (ions or neutrals) is much larger than the average distance between grains. Hence, at these spatial scales
particles move along ballistic trajectories until they hit the surface of a grain (for simplicity, the depicted tracks are
straight). Due to the presence of the other grain(s), particles collide less frequently with a part of the surface in the
shadow zone (between the grains). This exerts a force on the left grain toward the right grain, and vice versa.}
\label{shadow}
\end{figure}

Recently it has been theoretically shown \citep[][]{Tsytovich2014} that interstellar dust is intrinsically unstable against
the formation of compact clouds whose size may vary from fractions to dozens of AU. The clouds are formed due to the
so-called ``shadowing'' forces that tend to bind dust together, as illustrated in Figure~\ref{shadow}. These forces, exerted
between individual grains by surrounding ions and neutrals, represent a generic mechanism of attractive collective
interactions in weakly ionized gases \citep[][]{Khrapak2001}. The equilibrium density in a cloud is reached when the
shadowing is balanced by the mutual electrostatic repulsion of grains (charged due to a combined effect of photoemission and
collection of electrons and ions). It was demonstrated that such clouds should produce gaseous clumps with the local density
much larger than the density of ambient ISM.

In this paper we study a possible influence of the compact dusty clouds on the rate of H$\to$H$_2$ transition in diffuse
ISM. In Section~\ref{formation} we summarize the mechanisms leading to the formation of dusty clouds and the creation of
gaseous clumps around them. In Section~\ref{H2 formation} we discuss the main processes determining the efficiency of the
recombination (association) of atomic hydrogen on dust, and estimate the expected average enhancement of the H$_2$
formation. In Section~\ref{conclusion} we mention various mechanisms that have been proposed previously to explain available
observational data, indicating the existence of tiny-scale gaseous clumps in diffuse ISM, and argue that the mechanism
associated with the compact dusty clouds provides optimum conditions for H$\to$H$_2$ transition. Finally, we point out some
astrophysical implications of the enhanced H$_2$ formation due to the presence of dusty clouds.

\section{Dusty clouds and gaseous clumps}
\label{formation}

\begin{figure*}\centering
\includegraphics[width=.8\textwidth,clip,trim=0.cm 5.3cm 0.cm 5.3cm,]{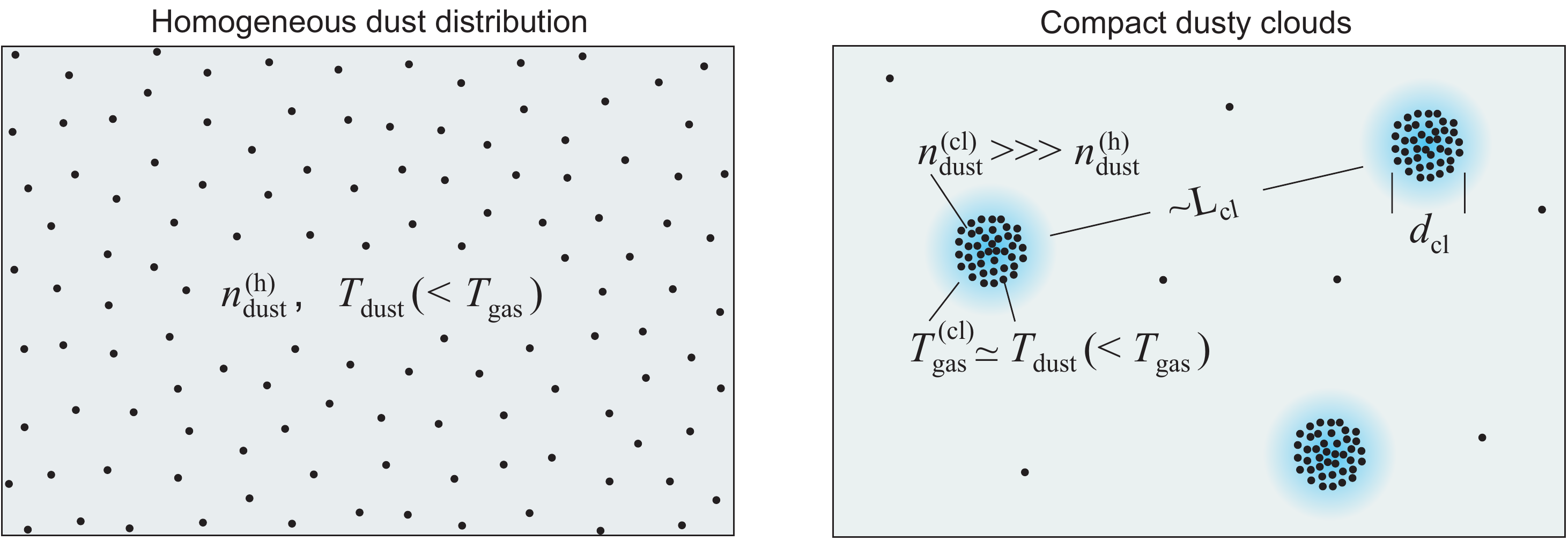}
\caption{A tiny-scale structure of the ISM, associated with compact dusty clouds \citep[not to scale, see][]{Tsytovich2014}.
A homogeneous dust distribution is intrinsically unstable against the formation of compact spherical clouds in diffuse ISM,
where the surface temperature of dust, $T_{\rm dust}$, is usually significantly smaller than the gas temperature $T_{\rm
gas}$. The dust density inside the clouds, $n_{\rm dust}^{\rm(cl)}$, is drastically increased with respect to the density
$n_{\rm dust}^{\rm(h)}$ in a homogeneous case, the resulting average distance between the clouds, $L_{\rm cl}$, is
$\sim\sqrt[3]{n_{\rm dust}^{\rm(cl)}/n_{\rm dust}^{\rm(h)}}$ times their size $d_{\rm cl}$.  Efficient radiative cooling
ensures that the dust surface temperature is the same in both cases (and equal to $T_{\rm dust}$). The gas temperature {\it
inside} the clouds, $T_{\rm gas}^{\rm(cl)}$, is close to $T_{\rm dust}$ due to strong thermal coupling; {\it outside} the
clouds the gas temperature rapidly tends to $T_{\rm gas}$ (same as in a homogeneous case). Therefore, a gaseous ``clump'' is
formed around each cloud (indicated by the blue shading), where the local gas density is increased by a large factor of
$T_{\rm gas}/T_{\rm dust}$ with respect to the ambient gas density. While the dust is virtually absent between the clouds,
the gas density in this region remains practically unchanged compared to a homogeneous case. Typical values of $d_{\rm cl}$
in diffuse ISM vary from fractions to dozens of AU.} \label{clouds}
\end{figure*}

Compact dusty clouds have been proposed by \citet[][]{Tsytovich2014} to form in diffuse ISM due to attractive shadowing
forces -- generic collective interactions between dust grains in a gaseous environment \citep[][]{Khrapak2001}. Both, the
ionized and neutral gas components contribute to the shadowing: Ions, recombining with electrons on the grain surface,
always result in attractive interactions, while the sign of the intergrain forces exerted by atoms depends on the
temperature difference: the interactions are attractive when the gas temperature $T_{\rm gas}$ is higher that the grain
surface temperature $T_{\rm dust}$, and are repulsive otherwise \citep[][]{Khrapak2001}. In diffuse ISM regions, such as the
cold or warm neutral medium (CNM or WNM), where $T_{\rm gas}$ is in the range between $\simeq100$~K and a few thousands of K
\citep[e.g.,][]{Draine2011Book} while $T_{\rm dust}$ does not exceed $\simeq20$~K \citep[e.g.,][]{Meisner2015}, the neutral
shadowing is thus attractive, too.

The shadowing forces compress dust into compact clouds, as sketched in Figure~\ref{clouds}, where the equilibrium is reached
due to the mutual electrostatic repulsion between charged grains. The dust density inside a cloud, $n_{\rm dust}^{\rm(cl)}$,
can be many orders of magnitude higher than the average dust density $n_{\rm dust}^{\rm(h)}$ in a homogeneous case: e.g.,
the maximum value of $n_{\rm dust}^{\rm(cl)}/n_{\rm dust}^{\rm(h)}$ for $\sim0.03~\mu$m grains is predicted to be as high as
$\sim10^7$ \citep[][this factor only slightly depends on the grain size]{Tsytovich2014}. Despite such a high local density,
the dust coagulation inside the clouds is practically inhibited because of a strong electrostatic repulsion between grains
(whose charges $Z$ in diffuse ISM regions are usually positive due to photoemission by the interstellar UV
field).\footnote{According to \citet[][]{Draine2011Book}, grains larger than $\sim 10^{-2}~\mu$m are always multiply charged
in diffuse ISM; the surface potential $eZ/a$ of a grain of radius $a$ can be roughly estimated as $\sim1-2$~V (WNM) or
$\sim0.3$~V (CNM). The resulting electrostatic barrier $e^2Z^2/a$ exceeds the thermal kinetic energy $k_{\rm B}T_{\rm gas}$
by more than one (WNM) or two (CNM) orders of magnitude.} Extinction in dusty clouds is governed by the Rayleigh regime, and
is very weak in the visible spectral range \citep[][]{Tsytovich2014}. It may be significant in the UV, in a cloud formed by
submicron dust \citep[representing the upper-size bound of the MRN-like distributions, e.g.,][]{Kim1994,Weingartner2001b}.
However, for small (e.g., $\sim 10^{-2}~\mu$m) grains, extinction in dusty clouds is negligible.

Dense dusty clouds can, in principle, be distorted by turbulent motion present in the ISM down to sub-AU length scales
\citep[e.g.,][and references therein]{Draine2011Book}. Nevertheless, the generic shadowing mechanisms of dust compression
remain unaffected by subsonic velocity perturbations. Since the detected relative fluctuations of the electron density,
associated with turbulence at such small scales, are of the order of $10^{-3}-10^{-4}$ \citep[][]{Cordes1985}, the
corresponding velocity perturbations must be very subsonic. Therefore in this paper we assume equilibrium properties of the
clouds, as derived by \citet[][]{Tsytovich2014} for a quiescent environment.

\subsection{Creation of gaseous clumps}
\label{gas_clumps}

One of the features of dusty clouds, playing -- as shown in the next section -- the crucial role for the surface chemistry,
is an efficient thermal coupling between dust and gas. The coupling is strong because the equilibrium cloud size is
self-regulated to be of the order of (or much larger than) the local mean free path of neutrals, which is primarily
determined by their collisions with grains \citep[][]{Tsytovich2014}. The right panel of Figure~\ref{clouds} shows the
effect on the gas distribution: Due to thermal accommodation of atoms colliding with dust, gas inside the clouds acquires
the temperature of the dust surface, $T_{\rm gas}^{\rm(cl)}\simeq T_{\rm dust}$. As a result, the gas density inside the
clouds increases by a factor of $\simeq T_{\rm gas}/T_{\rm dust}$ with respect to the ambient density,\footnote{The gas
temperature outside the clouds is the same as in the homogeneous case; we denote it for brevity by $T_{\rm gas}$.} to ensure
a constant gas pressure, and thus each dusty cloud produces a local ``atmosphere'' -- a {\it gaseous clump}. The
conservation of the radial heat flux yields the profiles of the gas temperature and density outside the cloud, rapidly
approaching the ambient gas values \citep[][]{Tsytovich2014}. We note that the generation of clumps does not practically
affect the gas density in the dust-free regions, between the clouds: Taking into account that the average distance between
the clouds $L_{\rm cl}$ and their characteristic size $d_{\rm cl}$ are related as $(L_{\rm cl}/d_{\rm cl})^3\sim n_{\rm
dust}^{\rm(cl)}/n_{\rm dust}^{\rm(h)}$ and using the global gas conservation, we obtain that the resulting relative decrease
of the gas density (with respect to a homogeneous case) is of the order of $(n_{\rm dust}^{\rm(h)}/n_{\rm
dust}^{\rm(cl)})(T_{\rm gas}/T_{\rm dust})$, i.e., is very small for all reasonable values of the temperature contrast.

The creation of clumps occurs at a timescale of the gas diffusion, which is equal to the squared size of a dusty cloud
divided by the diffusion coefficient of gas particles. The latter approximately equals to $v_{\rm H}/\sigma n_{\rm H}$
\citep[e.g.,][]{Smirnov2006book}, where $v_{\rm H}=\sqrt{k_{\rm B}T_{\rm gas}/m_{\rm H}}$ is the thermal velocity scale of H
atoms, $\sigma\sim10^{-15}$~cm$^2$ is the gas-kinetic cross section of their mutual collisions, and $n_{\rm H}$ is their
density. Thus, the characteristic time of the clump creation is of the order of $\sim d_{\rm cl}^2\sigma n_{\rm H}/v_{\rm
H}$. For typical values of $n_{\rm H}\sim 10^2$~cm$^{-3}$ and $v_{\rm H}\sim10^5$~cm/s this process is very fast, taking a
few yr for a 1~AU cloud \citep[][]{Tsytovich2014}. A timescale of the formation of dusty clouds is a factor of $\sim
\sqrt{m_{\rm dust}/m_{\rm H}Z}$ longer, where $m_{\rm dust}$ is the grain mass and $Z$ is the charge number: e.g., for
$\sim0.03~\mu$m grains with $m_{\rm dust}\sim3\times10^{-16}$~g and $Z\sim10$ dusty clouds are formed within $\sim10^5$~yr.
Still, this timescale is much shorter than the average lifetime of a diffuse molecular cloud (over 1~Myr).

It is noteworthy that the photoelectric (and cosmic-ray) heating and radiative cooling, whose balance governs the
equilibrium gas temperature in a homogeneous case, cannot noticeably affect the temperature inside the clouds: According to
\citet[][]{Draine2011Book}, the gas cooling function $\Lambda$ in diffuse ISM regions rapidly decreases with decreasing
$T_{\rm gas}$, and does not exceed a value of $\Lambda/n_{\rm H}^2\sim10^{-27}$~erg~cm$^3$~s$^{-1}$ for $T_{\rm
gas}\lesssim10^2$~K (this ratio is practically independent of $n_{\rm H}$). The resulting characteristic gas cooling time,
$n_{\rm H}k_{\rm B}T_{\rm gas}/\Lambda\propto n_{\rm H}^{-1}$, is longer than $\sim3\times10^2$~yr (for $n_{\rm H}\sim
10^2$~cm$^{-3}$ and $T_{\rm gas}\sim10^2$~K). This should be compared with the mean time of gas-dust collisions, $\sim(\pi
a^2 v_{\rm H}n_{\rm dust}^{\rm(cl)})^{-1}$, which also scales as $\propto n_{\rm H}^{-1}$. Setting $n_{\rm
dust}^{\rm(cl)}\sim10^7n_{\rm dust}^{\rm(h)}$ and estimating the characteristic density of ISM grains from the MNR size
distribution, $n_{\rm dust}^{\rm(h)}(a)\sim10^{-25}(a/{\rm cm})^{-2.5}n_{\rm H}$ \citep[][]{Weingartner2001b}, we obtain
that the collision time for $a\sim0.03~\mu$m is 2--3 orders of magnitude shorter than the cooling time. This justifies the
assumption of a perfect thermal coupling of gas to dust.

To conclude this section, we point out that the equilibrium cloud size is inversely proportional to the {\it local} gas
density \citep[][]{Tsytovich2014}. Thus, the creation of a gaseous clump should stimulate a breakup of the original dusty
cloud into smaller equilibrium clouds whose characteristic size is a factor of $\simeq T_{\rm gas}/T_{\rm dust}$ smaller
than the original size, while the local gas density inside these smaller clouds remains unchanged (and equal to $\simeq
T_{\rm gas}/T_{\rm dust}$ times the ambient gas density). While the breakup kinetics, not considered by
\citet[][]{Tsytovich2014}, is an interesting problem by itself, it does not qualitatively affect the process of H$_2$
formation, which is the main focus of the present work. Therefore we leave the analysis of this problem for a separate
paper.

\section{Formation of molecular hydrogen in diffuse ISM} \label{H2 formation}

The presence of dusty clouds makes all chemical reactions, occurring in diffuse ISM at the surface of grains, extremely
heterogeneous. Since the rates of the surface reactions are generally nonlinear functions of the gas density, it is natural
to expect that also the {\it global} rates (averaged over the tiny-scale inhomogeneities) must be affected by the presence
of dusty clouds.

Let us elaborate on the effect of dusty clouds on the formation of H$_2$. The equilibrium molecular density in the gas phase
is determined by the balance between the photodissociation, which is the principal process destroying interstellar H$_2$,
and the atomic recombination at the grain surface \citep[][]{Jura1974,Jura1975,Draine2011Book,Wakelam2017}. While the
photodissociation term in the corresponding balance equation is linearly proportional to the molecular density (the same is
true for other relevant dissociation processes, e.g., due to CRs), the recombination term is essentially nonlinear.

In the framework of the basic Langmuir kinetics for the interaction of hydrogen atoms with dust, the equilibrium surface
coverage of the physisorbed sites $\varphi$, equal to the areal density of hydrogen atoms multiplied by the site area $S$
(for simplicity we assume $\varphi\ll1$), is described by the following balance equation
\citep[e.g.,][]{Biham2002,Wakelam2017}:
\begin{equation}\label{rate}
Sj_{\rm H}\simeq\varphi \nu \exp\left(-\frac{E_{\rm des}}{k_{\rm B}T_{\rm dust}}\right)
+2\varphi^2\nu \exp\left(-\frac{E_{\rm diff}}{k_{\rm B}T_{\rm dust}}\right).
\end{equation}
The lhs is determined by the effective flux of incoming hydrogen atoms per unit area, $j_{\rm H}=\frac1{\sqrt{2\pi}}sn_{\rm
H}v_{\rm H}$ (i.e., $Sj_{\rm H}$ is the effective flux in units of monolayers/s), which is proportional to the sticking
probability $s$, a function of the gas temperature. Since typical $T_{\rm dust}$ in diffuse ISM is about 15~K
\citep[][]{Tielens2005book,Hocuk2017}, and the sticking probability in chemisorbed sites for {\it local} gas temperatures
($\simeq T_{\rm dust}$) is extremely low \citep[][]{Cazaux2011}, the physisorption is assumed to be the main adsorption
mechanism. The first depopulation term on the rhs represents the thermally activated desorption of atoms, the second term is
due to their recombination induced by thermally activated diffusion, where $E_{\rm des}$ and $E_{\rm diff}$ are the
respective activation energy barriers and $\nu$ is the typical attempt rate characterizing thermal hopping of the adsorbed
atoms.

The diffusion energy comprises a certain fraction of the desorption energy, $E_{\rm diff}=\alpha E_{\rm des}$
\citep[][]{Wakelam2017}. The exact value of $\alpha$ is unknown, and for different atoms it may vary from $\simeq0.3$
\citep[][]{Hasegawa_ea1992} to $\simeq0.7$ \citep[][]{Minissale2016}. The magnitude of $E_{\rm des}/k_{\rm B}$ is estimated
to be about 370~K for olivine and about 660~K for amorphous carbon \citep[e.g.,][]{Katz1999}. This implies that both
depopulation terms on the rhs of Equation~(\ref{rate}) have a very sharp dependence on the dust temperature. The desorption
and recombination terms dominate the depopulation at higher and lower $T_{\rm dust}$, respectively, and a transition between
these two regimes is very sharp, too. Assuming $n_{\rm H}=10^2$~cm$^{-3}$ and $T_{\rm gas}=10^2$~K, which approximately
corresponds to H~I regions \citep[][]{Draine2011Book}, and setting $\nu\sim10^{12}~{\rm s}^{-1}$, from Equation~(\ref{rate})
we infer that the desorption dominates at $T_{\rm dust}\gtrsim13$~K for olivine and at $T_{\rm dust}\gtrsim23$~K for carbon
(for $\alpha=0.5$).\footnote{Since $E_{\rm diff}=\alpha E_{\rm des}\gg T_{\rm dust}$, from Equation~(\ref{rate}) it follows
that $\varphi$ is indeed very (exponentially) small in the desorption-dominated regime.} Therefore, in diffuse ISM the
surface kinetics of H atoms is expected to occur in the desorption-dominated regime for olivine grains, which represent the
main fraction of the interstellar dust \citep[][]{Hocuk2017}. For carbonaceous grains, on the other hand, the recombination
term should dominate the kinetics.

The rate of H$_2$ formation per unit dust area, described by the recombination term in Equation~(\ref{rate}), scales as
$\propto\varphi^2$. In the desorption-dominated regime, where $\varphi\propto j_{\rm H}$, this yields $R_{\rm H_2}\propto
j_{\rm H}^2n_{\rm dust}$ for the formation rate per unit volume, where $n_{\rm dust}$ is the relevant dust density. Note
that one can also add the Eley-Rideal (ER) recombination term $\sim \varphi Sj_{\rm H}$ \citep[][]{Cazaux2004} to the rhs of
Equation~(\ref{rate}), which does not change the obtained scaling dependence of $R_{\rm H_2}$ in this case. In the
recombination-dominated regime, where $\varphi^2\propto j_{\rm H}$, we obtain $R_{\rm H_2}\propto j_{\rm H}n_{\rm dust}$;
the ER term in this case is negligible as long as $\varphi\ll1$.

\subsection{Effect of dusty clouds}
\label{enhancement}

The simple consideration above allows us to draw an important general conclusion concerning the role of dusty clouds in the
average steady-state abundance of molecular hydrogen. The observationally-relevant (effective) characteristics of the ISM
should be derived by averaging over tiny-scale inhomogeneities introduced by the clouds, as $\langle\ldots\rangle=
\mathcal{V}^{-1} \int_{\mathcal{V}} \ldots dV$, where $\mathcal{V}\sim L_{\rm cl}^3$ is the ISM volume per dusty cloud (see
right panel of Figure~\ref{clouds}).

We point out that the characteristic timescale of H$_2$ formation in diffuse ISM is significantly longer than the diffusion
time of hydrogen at the intercloud distance $L_{\rm cl}$: For typical parameters $L_{\rm cl}\sim 10^2d_{\rm cl}\sim
10^2$~AU, $n_{\rm H}\sim10^2~$cm$^{-3}$, and $\sigma\sim10^{-15}$~cm$^2$, the diffusion time is $\sim L_{\rm cl}^2\sigma
n_{\rm H}/v_{\rm H} \sim3\times10^5$~yr, which is about two orders of magnitudes shorter than the H$_2$ formation timescale
for the same density \citep[][]{Draine2011Book}. Therefore, it is reasonable to assume that the molecular hydrogen generated
in the clumps is homogeneously distributed across the ISM volume due to diffusion.

In a H$_2$ balance equation, the (photo)dissociation term scales as $\propto n_{{\rm H}_2}$ and thus is obviously not
affected by the averaging. On the other hand, the integral over the formation rate $R_{\rm H_2}\propto j_{\rm
H}^{2\beta}n_{\rm dust}$ is non-zero only within the cloud volume, where the dust density is $n_{\rm dust}^{\rm(cl)}$ and
$\beta$ varies between 1 (desorption-dominated regime, higher $T_{\rm dust}$) and 1/2 (recombination-dominated regime, lower
$T_{\rm dust}$). The value of $j_{\rm H}$ is proportional to the product of the local density of hydrogen atoms and the
square root of the local temperature. Since the gas pressure is constant across a clump, we have $j_{\rm H}\propto
\sqrt{n_{\rm H}}$. Setting the local gas density $n_{\rm H}$ inside dusty clouds equal to the ambient gas density multiplied
by a factor of $\simeq T_{\rm gas}/T_{\rm dust}$ and utilizing the global dust conservation, we derive the effective rate of
H$_2$ formation,
\begin{equation}\label{rec_rate}
\langle R_{\rm H_2}\rangle \simeq\left(\frac{T_{\rm gas}}{T_{\rm dust}}\right)^\beta R_{\rm H_2},
\end{equation}
where $R_{\rm H_2}$ is the formation rate corresponding to a homogeneous case (left panel of Figure~\ref{clouds}).

Thus, the relative enhancement of the global formation rate in the presence of dusty clouds (and, hence, the increase of the
molecular abundance in diffuse H~I regions) is expected to be about $(T_{\rm gas}/T_{\rm dust})^{\beta}$. For a typical dust
temperature about 15~K and gas temperatures varying between 80~K and 150~K toward different lines of sight
\citep[e.g.,][]{Falgarone2005,Snow2006, Gerin2015, Winkel2017}, one can expect that H$_2$ formation will be accelerated by a
factor of 5--10.

\section{Discussion and conclusion}
\label{conclusion}

There is a great deal of observational evidence indicating strong inhomogeneity of diffuse ISM at the length scales down to
$\sim$AU \citep[e.g.,][]{Dieter1976,Crawford2002,Lazio2009,Stanimirovic2010,Dutta2014,Corby2018}. Various mechanisms
proposed so far to explain this phenomenon include the formation of cold anisotropic (sheetlike) structures
\citep[][]{Heiles1997,Heiles2007}, fractal structures arising from MHD turbulence \citep[][]{Elmegreen1999} and
irregularities representing the tail-end of the turbulent spectrum \citep[][]{Deshpande2000}, structures generated due to
excitation of slow magnetosonic waves \citep[][]{Falle2002,Hartquist2003}, etc. However, only the mechanism associated with
the formation of compact dusty clouds enables creation of {\it equilibrium} gaseous clumps. This latter feature is
particularly important for the H$_2$ formation, providing optimum conditions for the heterogeneous enhancement of the atomic
recombination discussed in Section~\ref{enhancement}.


The results of Section~\ref{enhancement} show that the existence of gaseous clumps around dusty clouds in diffuse ISM may
shorten the timescale of H$\to$H$_2$ transition by up to an order of magnitude, thus reducing it to a few Myr in our model.
The timescale of H$_2$ formation has a broad astrophysical importance -- in particular, it determines the possible physical
scenarios of evolution of giant molecular clouds (GMCs). It has been suggested that long timescales (exceeding
$\sim10^7$~yr) favor the scenario of slow evolution of typical GMCs \citep[with $n_{\rm H}\sim10^2$~cm$^{-3}$,][]{Blitz1980}
as gravitationally bound objects in virial equilibrium, supported either by magnetic fields or by turbulence
(Tassis\&Mouschovias 2004, Mouschovias et al. 2006, Matzner 2002, Krumholz et al. 2006).

An alternative theory of formation of GMCs implies that they are the transient objects in diffuse medium, formed and
destroyed by large-scale turbulent flows on timescales of $\sim$~Myr \citep[e.g.,][]{MacLow2004}. This is inconsistent with
the timescales of H$_2$ formation in a homogeneous quiescent gas \citep[][]{Hollenbach1971b}. Furthermore, there are
indications that the lifetime of molecular clouds may indeed be relatively short -- these include, e.g., the lack of post-T
Tauri stars older than 3 Myr in nearby star-forming regions and the absence of mechanisms that can prevent fast
gravitational collapse and fragmentation of newly-formed clouds \citep[][]{Hartmann2001}.\footnote{Note, however, that a
rapid star formation scenario proposed by \citet[][]{Hartmann2001} for the Taurus-Auriga association is debated by
\citet[][]{Palla2002}.} Also, the high fraction of molecular hydrogen observed in turbulent diffuse ISM implies H$_2$
formation timescales on the order of $10^6$~yr \citep[see][and references therein]{Liszt2007}.

To the best of our knowledge, all existing scenarios of rapid H$\to$H$_2$ transition in diffuse ISM and formation of GMCs
assume inhomogeneities in gas and dust densities caused by dynamical processes, such as cloud collisions or turbulence on
different scales \citep[e.g.,][]{Glover2007,Lesaffre2007,Godard2009,Micic2012,Valdivia2016}. The possible existence of
equilibrium gaseous clumps, considered in this work, leads to a new type of physical mechanism that can significantly
accelerate the H$_2$ formation in diffuse medium and foster the efficient H$\to$H$_2$ transition. Obviously, chemistry of
other species in diffuse ISM is also affected by possible presence of such clumps, but we leave this question for a separate
study.

\section*{Acknowledgements}
This work was partially supported by the Russian Science Foundation (project 18-12-00351).


\end{document}